\documentclass[apjl]{emulateapj}

\slugcomment{To appear in ApJL}

\usepackage{epsfig} 
\usepackage{graphicx}
\begin{document}

\title{Discovery of an expanding molecular bubble in Orion BN/KL}

\shortauthors{Zapata, et al.}

\author{Luis A. Zapata\altaffilmark{1,2}, Laurent Loinard\altaffilmark{1},
Johannes Schmid-Burgk\altaffilmark{2}, Luis F. Rodr\'\i guez\altaffilmark{1},\\
Paul T. P. Ho\altaffilmark{3,4}, and Nimesh A. Patel\altaffilmark{3}}
 \altaffiltext{1}{Centro de
  Radioastronom\'\i a y Astrof\'\i sica, UNAM, Apdo. Postal 3-72
  (Xangari), 58089 Morelia, Michoac\'an, M\'exico} 
 \altaffiltext{2}{Max-Planck-Institut
  f\"{u}r Radioastronomie, Auf dem H\"ugel 69, 53121, Bonn, Germany}
 \altaffiltext{3}{Harvard-Smithsonian Center for Astrophysics, 
60 Garden Street, Cambridge, MA 02138, USA}
\altaffiltext{4}{Academia Sinica Institute of Astronomy and Astrophysics,
Taipei, Taiwan}

\begin{abstract} 
 During their infancy, stars are well known to expel matter violently in the form of 
 well-defined, collimated outflows. A fairly unique exception is found in the Orion 
 BN/KL star-forming region where a poorly collimated and somewhat disordered 
 outflow composed of numerous elongated ``finger-like'' structures was discovered
 more than 30 years ago. 
 In this letter, we report the discovery in the 
 same region of an even more atypical outflow phenomenon. Using $^{13}$CO(2-1) line 
 observations made with the Submillimeter Array (SMA), we have identified 
 there a 500 to 1,000 years old, expanding, roughly 
 spherically symmetric bubble whose characteristics are entirely different from those 
 of known outflows associated with young stellar objects. The center of the bubble 
 coincides with the initial position of a now defunct massive multiple stellar system 
 suspected to have disintegrated 500 years ago, and with the center of 
 symmetry of the system of molecular fingers surrounding the Kleinmann-Low 
 nebula. We hypothesize that the bubble is made up of gas and dust that 
 used to be part of the circumstellar material associated with 
 the decayed multiple system.
The Orion hot core, recently proposed to be the result of the impact of a shock
 wave onto a massive dense core, is located toward the south-east quadrant 
 of the bubble. The supersonic expansion of the bubble, and/or the impact of some 
 low-velocity filaments provide a natural explanation for its origin.
\end{abstract}

\keywords{stars: formation -- ISM: jets and outflows -- techniques: imaging spectroscopy
   -- techniques: high angular resolution -- ISM: molecules}

\section{Introduction}

Discovered more than 40 years ago, the Becklin-Neugebauer/Kleinmann-Low (BN/KL) 
region in Orion \citep{Beck1967,Klei1967} remains quite enigmatic. 
It is at the center of a fast  (30--100 km 
s$^{-1}$), and massive ($\sim$ 10 M$_{\odot}$) outflow particularly prominent in molecular 
hydrogen emission at near-infrared wavelengths \citep{kwan1976,allen1993,Taylor1984}. 
Composed of numerous
elongated ``finger-like'' structures, this outflow has a very peculiar morphology suggestive 
of an explosive origin. A strong and turbulent hot molecular core \citep{Wri1996,Bla1996,Liu2002} 
associated with intense maser emission from several molecular species \citep{Gau1998,Cohen2006,zapa2009a} 
also lies in the BN/KL region, but the very nature of this hot core and its relation to the outflow remain debated. 
Three massive young stars (BN, {\it I} and {\it n}) are known to be moving 
away from the KL nebula \citep{plam1995,gomez2005,gomez2008} at a few tens of km s$^{-1}$. 
Their velocity vectors 
point away from a common point of origin where the three stars must have been located 
about 500 years ago. This suggests that they were originally members of a multiple system 
that dynamically decayed very recently \citep{rod2005,gomez2005,gomez2008}.  
Remarkably, our recent molecular 
observations of the outflow associated with BN/KL \citep{zapa2009b} showed that the elongated 
fingers forming that flow also point exactly away from the position where the initial multiple 
stellar system was located. Moreover, in this study, we found that the dynamical age of 
the outflow  is consistent with about 500 years. It is, therefore, very likely that the same 
energetic phenomenon that led to the acceleration of the stellar sources BN, {\it I} and 
{\it n} was also responsible for the explosion that gave birth to the peculiar gaseous 
outflow in the BN/KL region of Orion.

\section{Observations}

In order to further  explore the interrelation between the various energetic phenomena
occurring in Orion BN/KL, we have carried out interferometric observations of the 
region in the optically thin 2$\rightarrow$1 line of $^{13}$CO at $\nu$ = 220.3 GHz using 
the Submillimeter Array (SMA) \citep{Ho2004}. This transition is a reliable indicator of the kinematics
and mass of the region observed. The observations were collected in 2007 January 
and 2009 February while the SMA was in its compact and sub-compact configurations, with 
baselines ranging in projected length from 6 to 58 m; this provides an angular resolution 
of about 4$''$. The digital correlator was configured to provide a spectral resolution of 0.40 
MHz, corresponding to a resolution in velocity of 1.05 km s$^{-1}$. The zenith opacity 
($\tau_{230GHz}$) was $\sim$ 0.1--0.3, indicating reasonable weather conditions. 
Observations of Uranus and Titan provided the absolute scale for the flux density calibration. 
The gain calibrators were the quasars 0530$+$135, 0541$-$056, and 0607$-$085. 

The data were calibrated using the IDL superset MIR, originally
developed for the Owens Valley Radio Observatory \citep{sco1993}
and adapted for the SMA\footnote{https://www.cfa.harvard.edu/$\sim$cqi/mircook.html}. 
The calibrated data were imaged
and analyzed in a standard manner using the MIRIAD, GILDAS, and AIPS
packages. We used the ROBUST parameter set to 2 to obtain a slightly better
sensitivity sacrificing some angular resolution. The line image
rms noise was around 100 mJy beam$^{-1}$ for each channel.

\begin{figure}[!t]
\begin{center} 
\includegraphics[scale=0.38, angle=0]{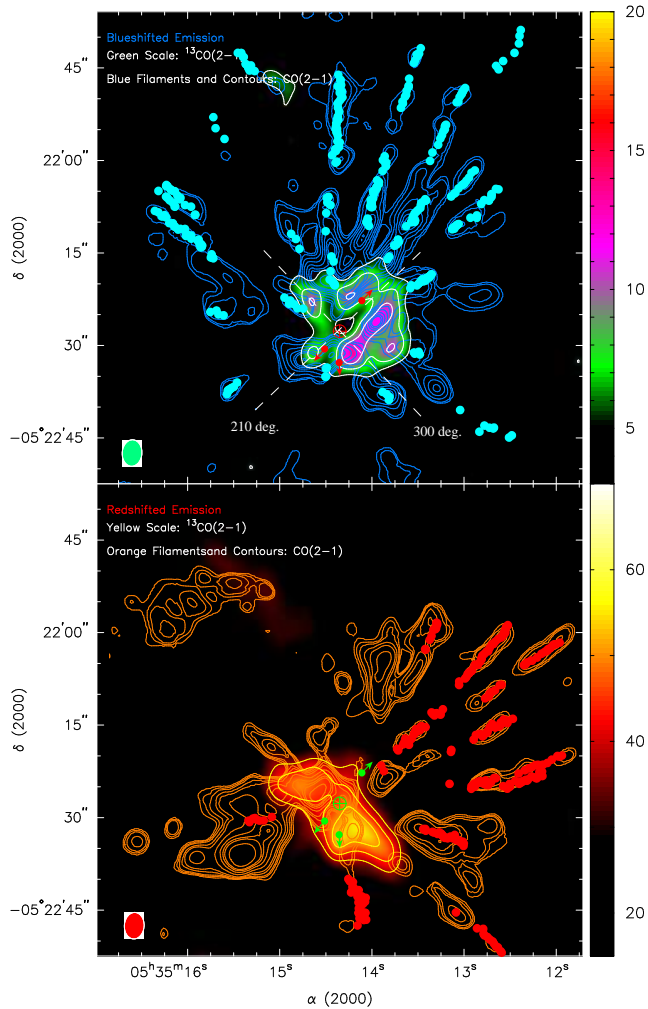}
\caption{\scriptsize Color and contour images of the blueshifted (upper panel) and 
         redshifted (lower panel) integrated $^{13}$CO(2-1) line emission from 
         the molecular bubble overlaid with the $^{12}$CO(2-1) filaments. 
         {\it Upper panel:} The white contours represent the blueshifted $^{13}$CO(2-1) 
          emission and are at $-$2, 2, 3, and 4 times 2.5 Jy beam$^{-1}$ km s$^{-1}$. 
         The blue contours represent blueshifted $^{12}$CO(2-1) emission and
         are at $-$4, 4, 8, 16, 24, 32, 38, 42, and 52 times 2 Jy beam$^{-1}$ km s$^{-1}$.
         {\it Lower panel:} The yellow contours represent the redshifted $^{13}$CO(2-1) emission 
         and are at $-$2, 2, 3, 4 and 5 times 13 Jy beam$^{-1}$ km s$^{-1}$. The orange contours 
         represent redshifted $^{12}$CO(2-1) emission 
         and are at $-$4, 4, 8, 16, 24, 32, 38, 42, and 52 times 2.7 Jy beam$^{-1}$ km s$^{-1}$.
         In both panels the blue and red points represent the position of the 
         $^{12}$CO(2-1) emission peaks in every velocity channel.  
         The synthesized beam of the line image is 4\farcs3 $\times$ 3\farcs2 
         with a P.A. of -4.0$^\circ$ and is shown near the lower left corner of each panel.
         The wedge indicates the line emission in mJy beam$^{-1}$ km s$^{-1}$.
         The red and green filled dots mark the positions of the runaway sources {\it BN}, {\it I}
         and {\it n} \citep{gomez2005}. The vectors on these sources represent the direction 
         of their proper motion \citep{rod2005,gomez2005,gomez2008}. The red
         and green cross-hair circles marks the zone from where the three sources were ejected 
         some 500 years ago \citep{gomez2005}  and the origin of the explosive flow \citep{zapa2009b}.
         The dashed lines in the top panel mark the position and orientation of the position-velocity 
         diagrams presented in Figure 2 were computed. }
\end{center}
\end{figure}

\begin{figure}[!t]
\begin{center} 
\includegraphics[scale=0.36, angle=0]{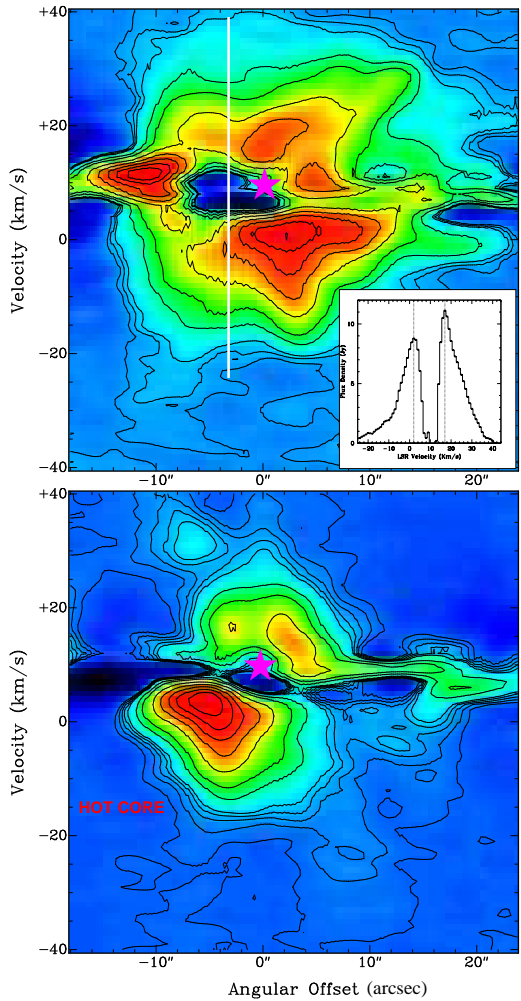}
\caption{\scriptsize $^{13}$CO(2-1) position-velocity digrams of the molecular ring computed 
at a position angle of $+$210$^\circ$ (upper panel) and $+$300$^\circ$ (lower panel). 
 The contours in both panels are from 26\%, 27\%, 29\%, 30\%, 40\%, 50\%, 60\%, 70\%, 80\% and 90\%
of the peak of the line emission. The systemic LSR radial velocity 
of the ambient molecular cloud is about 9 km s$^{-1}$. The synthesized beam of 
the line image is 4\farcs3 $\times$ 3\farcs2 with a P.A. of $-$4.0$^\circ$. 
The pink star marks the position and LSR velocity of the center of the
explosive outflow \citep{zapa2009b}. The emission at ambient velocities is 
clearly extended and poorly sampled with the SMA.  In the right bottom corner
of the upper panel, we show a intensity-velocity cut of the ring at a P.A. = $+$90$^\circ$ 
(see the white line) which reveals that expansion velocity of bubble is approximately 15 km s$^{-1}$.  
In the bottom panel, we mark the position of the Orion Hot Core. }
\end{center}
\end{figure}

\section{Results and Dicussion}

The $^{13}$CO(2-1) line detected in our observations is broad, extending from about $-$30 
km s$^{-1}$ to $+$40 km s$^{-1}$ in radial velocity. The structure of the emitting region 
is shown in Figure 1, where we present separately the blueshifted (from $-$31 to $-$18 km 
s$^{-1}$; upper panel) and redshifted ($+$23 to $+$33 km s$^{-1}$; lower panel) integrated 
intensity maps. In Figure 1, we also show (as contours) the integrated intensity of the high-velocity
fingers (from $-$130 to $-$35 km s$^{-1}$ for the approaching velocities, 
and from $+$35 to $+$120 km s$^{-1}$ for the receding velocities) traced by $^{12}$CO(2-1) 
emission \citep{zapa2009b}, and the positions and proper motions of the 
runaway sources ({BN}, {\it I} and {\it n}). The position of the multiple stellar system which
contained these sources before it decayed is shown as a crossed-haired circle, and we shall
henceforth refer to this point as ``O''. 

The $^{13}$CO(2-1) emission clearly traces a roundish structure curling around position ``O''. The 
emission at position ``O'' itself is weak so the structure must be hollow: it is either a ring or 
a shell. In Figure 2, we show the $^{13}$CO(2-1) position-velocity diagrams of that structure 
computed along strips centered on ``O'' along position angles of $+$210$^\circ$ and $+$300$^\circ$.
The morphology seen in these diagrams is characteristic of a bubble: redshifted and blueshifted
emission are simultaneously present toward the center of the structure, corresponding to the
two sides of the bubble. It is impossible from its morphology and kinematics alone to decide 
whether the bubble is expanding or contracting. The fact that the three stars (BN, {\it I}, and 
{\it n}) known to be moving away from point ``O'' are projected exactly on top of the bubble, 
however, almost certainly indicates that it is expanding. The expansion velocity of the bubble 
is of order 15 km s$^{-1}$ (Figure 2) and, given the distance to Orion \citep[414 pc;][]{men2007}, 
its radius is about 2000 AU. This yields a dynamical age for the bubble of about 600 years, with an
uncertainty of a few hundred years. 

The C$^{18}$O(2-1) line was also included in our SMA observations. The bubble is marginally
seen in these data, with a typical intensity 6 to 8 times weaker than in the $^{13}$CO(2-1) line.
This indicates that the $^{13}$CO(2-1) line is optically thin, and can be used to estimate the 
mass of the expanding bubble. Assuming an excitation temperature of 50 K, and an abundance
ratio between H$_2$ and $^{13}$CO of 5 $\times$ 10$^5$, we obtain a mass of about 5 
M$_\odot$ for the entire bubble. The corresponding kinetic energy (assuming an expansion 
velocity of 15 km s$^{-1}$) is about 1 $\times$ 10$^{46}$ erg, roughly one order of magnitude 
smaller than the kinetic energy of the outflow \citep{kwan1976}.  

The bubble is centered on point ``O'', which decidedly appears to be a very special position 
within the Orion BN/KL region. This, combined with the remarkably similar timescales of all
three phenomena, clearly indicates that the acceleration of the massive young stars BN, {\it I}, 
and {\it n}, and the births of both the fast outflow and the bubble most certainly occurred as a 
consequence of a single energetic event. Indeed, a natural scenario explaining the creation 
of the bubble follows from the interpretation of the large velocities of BN, {\it I}, and {\it n} in 
terms of a dynamical disintegration. Young stars are well-known to be surrounded by 
circumstellar disks and envelopes. If a multiple system disintegrates because of a chaotic 
few-body encounter, the circumstellar material initially in those structures will clearly be 
affected. The gas most strongly bound to the young stars may be ejected together with the 
stars themselves, and some of the material may even be independently ejected to high 
speed by a sling-shot effect. The least strongly bound material, however, will likely remain 
nearly stationary at first. However, since the stars that provided most of the gravitational 
binding energy of the young system are now gone, this material will find itself with an excess 
of kinetic energy, and will start to expand at a speed typical of the velocity dispersion of
the material ($\approx$ 15 km s$^{-1}$ as we will see momentarily). Moreover, since it was 
initially distributed among several disk/envelope structures with randomly oriented angular 
velocities, this material would be expected to expand fairly isotropically rather that as a 
highly flattened system. Finally, because of conservation of angular momentum, the 
system will progressively loose any original flattening it might have had. 

As mentioned earlier, the kinetic energy of the bubble is about 1 $\times$ 10$^{46}$ erg,
a factor of 100 to 1000 times less than the disruptive energy of {\it I}, and {\it n} and BN 500 
years ago \citep{Bally2005}. Its 
gravitational energy, on the other hand, is about $-{GM^2 \over R}$ = $-$ 2 $\times$ 10$^{44}$ 
erg. Thus, the bubble is clearly not currently  in virial equilibrium, and must be expanding --in 
agreement with our previous considerations. Before the disintegration, however, BN, {\em I}, 
and {\em n} as well as the 15 $M_\odot$ in the outflow and bubble were all part of a common 
system. The combined mass of BN, {\em I}, and {\em n} must be of order 35 $M_\odot$, so the
total mass of the system would have been about 50 $M_\odot$. For such a system to be in
virial equilibrium, it must be confined to a radius of about 200 AU. Interestingly, this is approximately
the size of the region (1$''$) where the sources BN, {\em I}, and {\em n} 
were located about 500 years ago \citep{gomez2005} . Moreover, the typical speed
of the gas in such a virialized structure would be $v$ = $\sqrt{GM \over R}$ $\approx$
15 km s$^{-1}$, in excellent agreement with the measured expansion velocity of the
bubble.

The expansion velocity of the bubble is clearly supersonic, so shocks should be created
where it encounters dense gaseous clumps during its expansion. Interestingly, our SMA 
observations do reveal the presence of typical shock tracers ({\it e.g.} SiO, SO, and 
SO$_2$) in the bubble. The Orion hot core mentioned earlier is located toward the 
south-east quadrant of the bubble (see Figure 2), and was recently suggested to be 
caused by the impact of a shock wave related to the dynamical decay of BN, {\it I}, 
and {\it n} onto a dense pre-existing core called the ``Extended Ridge'' \citep{zapa2010}. 
To explain the near systemic mean velocity of the hot core, the authors of that suggestion 
argued that the shock wave must be fairly slow, and interact with a particularly dense and 
massive material.  \citep{zapa2010} proposed that the required shock wave might be due
to the impact of low-velocity filaments. %It is noteworthy, however, that there are very few
%filaments in the direction of the hot core (see Figure 1). 
However, the supersonic expansion of the
bubble reported here might provide a more natural explanation of the shocks required 
to explain the Orion Hot Core. For a more complete discussion of the nature of the
Orion Hot Core, we refer the readers to \citet{zapa2010}. 

\section{Conclusions}
In this letter, we reported the discovery of an entirely new type of outflow phenomenon 
in the BN/KL region of Orion. Using $^{13}$CO(2-1) SMA observations, we identified a 500 to 1000  
years old, roughly spherically symmetric bubble, expanding at about 15 km s$^{-1}$.
The center of the bubble coincides with the initial position of a now defunct massive 
multiple stellar system suspected to have disintegrated 500 years ago, and with the 
center of symmetry of the system of molecular fingers surrounding the Kleinmann-Low 
nebula. We propose that the material in that bubble was originally associated with 
the young stars in the now defunct multiple system.

\acknowledgments
L.A.Z., L.L. and L.F.R.\ acknowledge the financial support of DGAPA, UNAM and CONACyT, M\'exico. 
The Submillimeter Array is a joint project between the Smithsonian Astrophysical Observatory 
and the Academia Sinica Institute of Astronomy and Astrophysics and is funded by the Smithsonian 
Institution and the Academia Sinica.

\end{document}